# STATISTICS OF THE CHI-SQUARE TYPE, WITH APPLICATION TO THE ANALYSIS OF MULTIPLE TIME-SERIES POWER SPECTRA

P.A. Sturrock[1] and M.S. Wheatland[2]

[1]Center for Space Science and Astrophysics, Stanford University, Stanford, CA 94305-4060
[2]School of Physics, University of Sydney, Sydney, Australia

ABSTRACT

It is often necessary to compare the power spectra of two or more time series: one may, for instance, wish to estimate what the power spectrum of the combined data sets might have been, or one may wish to estimate the significance of a particular peak that shows up in two or more power spectra. Also, one may occasionally need to search for a complex of peaks in a single power spectrum, such as a fundamental and one or more harmonics, or a fundamental plus sidebands, etc. Visual inspection can be revealing, but it can also be misleading. This leads one to look for one or more ways of forming statistics, which readily lend themselves to significance estimation, from two or more power spectra. The familiar chi-square statistic provides a convenient mechanism for combining variables drawn from normal distributions, and one may generalize the chi-square statistic to be any function of any number of variables with arbitrary distributions. In dealing with power spectra, we are interested mainly in exponential distributions. One well-known statistic, formed from the sum of two or more variables with exponential distributions, satisfies the gamma distribution. We show that a transformation of this statistic has the convenient property that it has an exponential distribution. We introduce two additional statistics formed from two or more variables with exponential distributions. For certain investigations, we may wish to study the minimum power (as a function of frequency) drawn from two or more power spectra. In other investigations, it may be helpful to study the product of the powers. We give numerical examples and an example drawn from our solar-neutrino research.

1. INTRODUCTION

Power spectrum analysis is one of the most powerful tools for the study of time series. (See, for instance, Oppenheim, Schafer & Buck 1999.) There are many situations in which one would like to combine or compare the spectral properties or two or more time series. In some cases, it may be possible to actually combine the time series and carry out a power spectrum analysis of the combined data. However, this is not always possible, and it may not be convenient even when it is possible. In such cases, it may be more convenient to compare and contrast the power spectra derived from the different time series.

In our neutrino research, it has been necessary to compare power spectra obtained from different experiments (Caldwell & Sturrock 2003). The comparison of the time series themselves would, even if possible, be very difficult since two experiments may use different measurement techniques that give only indirect flux estimates and involve different time periods, different "run" times, different forms of contamination, different systematic effects, etc. It is important also to compare power spectra derived from neutrino experiments with power spectra formed from well-studied solar data such as X-ray emission (Sturrock & Weber 2002a). The comparison of the corresponding time series would again be very difficult.

In addition to looking for similarities between two different power spectra, one may need to determine whether there is a significant relationship between oscillations at two or more frequencies in the same power spectrum. For instance, we have found that the power spectrum formed from Homestake (Davis & Cox 1991; Lande et al. 1992; Cleveland et al. 1995, 1998) measurements shows not only a peak at 12.88 $y^{-1}$ due apparently to solar rotation, but also two sidebands close to 11.88 $y^{-1}$ and 13.88 $y^{-1}$ due apparently to the influence of the inclination to the ecliptic of the Sun's rotation axis (Sturrock, Walther, & Wheatland 1997; Sturrock, Walther, & Wheatland 1998). More recently, we have begun to investigate (Sturrock 2003a, 2003b) evidence for modulation of the neutrino flux by a certain group of oscillations (r-modes; see, for instance, Saio, 1982) that appear to be responsible for well-known oscillations similar to the Rieger oscillation (Rieger et al. 1984), with frequencies that are found to be sub-harmonics of a "fundamental" frequency (Bai 2003).

For these reasons, it is important to determine how much information one can obtain from power spectra alone. For the purpose of this article, it is generally assumed that the time-series from which the spectra were derived are unknown: the only relevant available information comprises the power spectra and the probability distribution functions of the powers for the null hypothesis that the time-series contain no systematic oscillations. In this context, we need to construct one or more procedures for determining whether two or more peaks in the same or different power spectra are related. These procedures should provide estimates of the significance of any such relationship.

It is well known that if a time series is formed from a sequence of measurements of a normally distributed random variable with mean zero and variance unity, the power S at any specified frequency $\nu$ is distributed exponentially:

$$P(S)dS = e^{-S}dS. \tag{1.1}$$

(See, for instance, Scargle 1982.) The probability of obtaining a power $S$ or more is given by the cumulative distribution function

$$C(S) = \int_S^\infty P(x)dx = e^{-S}. \tag{1.2}$$

Since S is a function of the frequency $\nu$, P and C also are expressible as functions of $\nu$. For some more complex problems, the power may be distributed in a more complex manner. Suppose that, on the basis of a null hypothesis, the power $\sigma$ is distributed according to the probability distribution function $P(\sigma)$ such that the probability that $\sigma$ lies in the range $\sigma$ to $\sigma + d\sigma$ is given by $P(\sigma)d\sigma$. If we introduce the statistic S defined by

$$S = -\ln(C) = -\ln\left(\int_\sigma^\infty d\sigma' P(\sigma')\right), \qquad (1.3)$$

we see that S is distributed exponentially. As such, S may be regarded as a proxy for the power or as a "normalized" power. It follows that we may, without loss of generality, restrict our attention in this article to measurements that are distributed exponentially.

The goal of this article is therefore to find one or more procedures for estimating the significance of two or more related measurements of variables that have exponential distributions. [Two of these procedures have previously been presented in a shorter article (Sturrock 2003c.] Strictly speaking, it is now irrelevant whether or not we are dealing with power spectra formed from time series. However, since the procedure has been developed primarily for application to power series, we shall refer to the relevant variables as "powers."

This problem is reminiscent of a standard problem in statistics, that of estimating the significance of two or more related measurements of variables which, if random, have normal distributions. A familiar and very useful technique is that of computing the chi-square statistic. (See, for instance, Bartoszynki & Niewiadomska-Bugaj 1996, p. 757, or Rice 1988, p. 168.) This suggests that we look for one or more similar statistics that are appropriate for the analysis of variables that have exponential distributions.

When using chi-square statistics, it is necessary to consult tables to determine the significance level of a particular estimate. In order to avoid this step, we plan to form statistics that have the following property: If the value of a statistic is found to be G, the significance level of this result is exactly the same as the significance level of a power spectrum that has the same value at a specified frequency. That is to say, the probability of obtaining the value G or more on the null (random noise) hypothesis would be $e^{-G}$.

In this article, the mathematical derivations are, as far as possible, relegated to appendices. This hopefully makes the article more readable and more useful. In Section 2, we introduce a statistic that is conceptually close to the chi-square statistic. We first form the sum of the variables (analogous to forming the sum of the squares of the variables in the context of normal distributions), and find the probability distribution function of that quantity. This is known as the "gamma distribution." (See, for instance, Bartoszynki & Niewiadomska-Bugaj 1996, p. 384, or Rice 1988, p. 39.) However, we then carry out a further operation to obtain a more useful function that has an exponential distribution. We refer to this quantity as the "combined spectrum statistic." In Section 3, we develop the "minimum power statistic," that is formed from the minimum (for each frequency value) of two or more power spectra. In Section 4, we introduce the "joint power statistic," that is formed from the product of two or more power spectra. In Section 5, we give the results of the application of these statistics to some simulated problems, and in Section 6 we give the results of the application of these methods to a problem we have met in our solar-neutrino research. Section 7 contains a brief discussion of the results of this article.

## 2. COMBINED POWER STATISTIC

If we wish to combine information from n power spectra, the combination that is closest to the chi-square statistic is the sum of the powers, which we write as

$$Z = S_1 + S_2 + ... + S_n. \qquad (2.1)$$

We show in Appendix B, that uses some general results derived in Appendix A, that the following function of Z, which we refer to as the "combined power statistic," is distributed exponentially:

$$G_n(Z) = Z - \ln\left(1 + Z + \tfrac{1}{2}Z^2 + ... + \tfrac{1}{(n-1)!}Z^{n-1}\right). \qquad (2.2)$$

Figure 1 gives plots of the combined power statistics of orders 2, 3, and 4.

## 3. MINIMUM POWER STATISTIC

We may wish to determine the frequency for which the minimum power among two or more power spectra has the maximum value. We therefore consider the following quantity, formed from the independent variables $x_1, x_2, \ldots$, each of which is distributed exponentially:

$$U(x_1, x_2, \ldots) = Min(x_1, x_2, \ldots) . \tag{3.1}$$

We show in Appendix C that the following function of U, that we refer to as the "minimum power statistic," is distributed exponentially:

$$K_n(U) = nU . \tag{3.2}$$

## 4. JOINT POWER STATISTICS

We next consider forming something resembling a "correlation function" by forming the product of two or more power spectra. We first consider just two power spectra, since this case lends itself to analytical treatment. It proves convenient to work with the square root of the product (the geometric mean), and we therefore write

$$X = (S_1 S_2)^{1/2} . \tag{4.1}$$

We show in Appendix D that the following function of X is distributed exponentially:

$$J_2 = -\ln(2X \, K_1(2X)), \tag{4.2}$$

where $K_1$ is the Bessel function of the second kind.

We show in Appendix F that we may determine simple functional fits to the joint power statistics. The fit to J2 is found to be

$$J_{2A} = \frac{1.943 X^2}{0.650 + X} . \tag{4.3}$$

We show J2 and J2A in Figure 2. The difference (mean and standard deviation) between the true value and the approximate value is only $0.019 \pm 0.036$ for 21 evaluations between X = 0 and X = 10. This difference is fortunately negligible, so there is no need to make calculations via Equation (4.2) or to consult a table.

We now consider joint power statistics of higher orders, and consider the following combination of n powers:

$$X = (S_1 ... S_n)^{1/n} . \tag{4.4}$$

For n > 2, it is not possible to find analytical functions of X that are distributed exponentially. However, we show in Appendix E how these functions may be computed. Using the procedure of Appendix F, we find that the joint power statistics of third and fourth orders are given to very good approximation by the following expressions:

$$J_{3A} = \frac{2.916 X^2}{1.022 + X} , \tag{4.5}$$

and

$$J_{4A} = \frac{3.881 X^2}{1.269 + X} . \tag{4.6}$$

These approximate expressions are compared with the calculated values in Figures 3 and 4, respectively. For 21 evaluations of X in the range 0 to 10, $J_{3A} - J_3 = 0.014 \pm 0.028$, with extrema –0.07 and 0.02, and $J_{4A} - J_4 = -0.002 \pm 0.020$, with extrema –0.03 and 0.05. We see that the approximate expressions are sufficiently accurate that there is no need to make further calculations or to consult tables.

## 5. SIMULATIONS

It is interesting to calculate the preceding statistics for simple numerical examples. By analogy with our power spectrum studies of the solar neutrino flux, we have formed synthetic spectra for frequencies in the range 0.01 to 40, in steps of 0.01. (In our neutrino studies, these would be frequencies measured in cycles per year.) For each frequency, a power $S_1$ was chosen from an exponential distribution using $S_1 = -\ln(u)$, where u is a uniformly distributed random variable in the range $0 < u < 1$. For the chosen signal frequency $\nu_S = 20$, a fixed power $P_s$ was inserted in the spectrum. This procedure was repeated to produce the four spectra $S_1(\nu),...,S_4(\nu)$. For the case $P_S = 5$, the spectra are shown in Figure 5. For this set of simulations, we have computed the three power statistics introduced in Sections 3, 4, and 5. A typical set of results is shown in Figure 6, where we see that the combined power statistic has the value 12.7 at the signal frequency [panel (a)]; the minimum power statistic has the value 20, of course [panel (b)]; and the joint power statistic has the value 15.5 [panel (c)]. We find that it is typically the case that if four spectra have similar values at the signal frequency, the minimum power statistic has the largest value and the combined power statistic the smallest value. However, this ordering does not hold for an arbitrary set of powers.

As a second numerical test of the statistics, we have considered the application to spectra produced from synthetic time series. For a sequence of $N$ unit-spaced times $t_i = i$ ($i = 0,1,...,N-1$), four time series

$$X_j(t_i) = A\sin(2\pi\nu_0 t_i) + R_{j,i} \tag{5.1}$$

were generated ($j = 1,2,3,4$), with a signal frequency $\nu_0 = 0.25$ per unit time, the noise term $R_{j,i}$ being generated from a normal distribution with zero mean and unit standard deviation. For each of these time series the classical periodogram (e.g. Scargle 1982)

$$S_j(\nu) = \frac{1}{N}\left|\sum_{k=0}^{N} X_j(t_k)\exp(-2\pi i k\nu)\right|^2 \tag{5.2}$$

was evaluated at the usual set of $N/2+1$ frequencies $\nu = \nu_l = l/N$ ($l = 0,1,2,...,N/2$), using the Fast Fourier Transform. Note that for the choice of unit spacing in the time series the highest frequency in this set (the Nyqvist frequency) is 0.5 per unit time, so the chosen signal frequency is half the Nyqvist frequency. The mean power at the signal frequency in the periodograms is $\bar{S}_0 = N(A/2)^2 + 1$ (Scargle 1982). We have chosen $\bar{S}_0 = 5$ and $N = 8192$, so that $A = 2[(\bar{S}_0 - 1)/N]^{1/2} \approx 0.044$. For the four spectra (periodograms) the combined power statistic, the minimum power statistic, and the joint power statistic were calculated at the signal frequency. This procedure was repeated 1000 times.

The results are summarized in Figure 7. Panel (a) shows the distribution of power at the signal frequency for the 4000 individual spectra (solid histogram). The spectra were constructed so that the mean of this distribution is $\bar{S}_0 = 5$, and this value is indicated by the dotted vertical line. The solid curve is the expected distribution $e^{-S}$ of noise, i.e. the distribution of power at frequencies other than the signal frequency, and this distribution is reproduced in each panel in the figure. Panel (b) shows the distribution of the value of the combined power statistic at the signal frequency (solid histogram). The mean of the observed distribution (indicated by the vertical dotted line) is about 12.7. Panel (c) shows the distribution of the minimum power statistic at the signal frequency (solid histogram); the mean of this distribution (dotted line) is about 9.1. Finally panel (d) shows the distribution of the joint power statistic at the signal frequency (solid histogram); the mean of this distribution (dotted line) is about 13.0. This figure illustrates how the three statistics lift the power at the signal frequency out of the noise, on average. In running these simulations, we have also verified that (at frequencies other than the signal frequency) the statistics are indeed distributed exponentially.

## 6. APPLICATION

As a typical example, we discuss briefly the application of these three statistics to an issue that arose in the course of our research concerning solar neutrino measurements.

In our analysis of Homestake data, we noticed that the main rotational modulation peak at 12.88 $y^{-1}$ seemed to be accompanied by "sidebands" near 11.88 and 13.88 $y^{-1}$, such as might be produced by a solar-latitude effect since the Sun's rotation axis is inclined with respect to the normal to the ecliptic (Sturrock, Walther, & Wheatland 1997, 1998). We have therefore computed the combined power statistic, the minimum power statistic, and the joint power statistic from $S_H(\nu-1)$, $S_H(\nu)$, and $S_H(\nu+1)$.

We show $S_H(\nu)$ and the three statistics in Figure 8. We see that each statistic clarifies the evidence that the peak at 12.88 is accompanied by two sidebands.

## 7. DISCUSSION

In preceding sections, we have introduced statistics formed from two or more power spectra that are distributed in the same way as simple power spectra formed from a time series of random variables with unit standard deviation. These statistics lead to estimates of the significance level of a peak found at a specified frequency. However, in practice, one will be searching for evidence of a peak in a specified band of frequencies. In our neutrino research, for instance, we may look for evidence for a peak in a band corresponding to the range of synodic rotation rates of the solar convection zone or the solar radiative zone. The standard technique for addressing this problem is to estimate the "false-alarm" probability (Press et al. 1992). We estimate (empirically or theoretically) the number N of independent frequencies in this band. Then the probability of obtaining a power S or larger by chance in this band is given by

$$P(\textit{false alarm}) = 1 - \left(1 - e^{-S}\right)^N . \qquad (7.1)$$

Since the statistics introduced in Sections 2, 3 and 4 also have exponential distributions, we may apply this formula to these statistics also. If this probability is small, it may be represented approximately by a "corrected" statistic (corrected to take account of the search band) given by

$$G^* \approx G - \ln(N) , \qquad (7.2)$$

with similar expressions for the statistics K and J.

However, it should be emphasized that it is always advisable to follow up significance estimates made in this way with a more robust test, such as the shuffle test. See, for example, Bahcall & Press (1991), Sturrock & Weber (2002b), and Walther (1999).

As we mentioned in Section 1, Bai (2003) has recently reviewed evidence that solar flares tend to exhibit periodicities with periods that are integer (2, 3,…) multiples of a "fundamental period" of approximately 25.5 days. The significance of this result could be assessed by means of a statistic formed from $S(\nu/2)$, $S(\nu/3)$, etc. Wolff (2002) has recently claimed to find somewhat similar patterns in power spectra formed from measurements of the solar radio flux. The statistics we have here introduced should be helpful in the further evaluation of such claims.

We have recently used these statistics in the analysis of Homestake and GALLEX-GNO solar neutrino data to seek and evaluate evidence of r-mode oscillations (see, for instance, Saio 1982) in these data (Sturrock 2003a, 2003b).

We are grateful to Jeffrey Scargle, Paul Switzer, and Guenther Walther for helpful discussions, and to Steve Harris for help with the analytical calculation of Section 4. This research was supported by NSF grant ATM-0097128.

APPENDIX A

We are interested in evaluating the significance of two or more measurements of variables that, if random, have known distributions. When the distributions are normal, one can adopt the chi-square statistic for this purpose. Since we are interested in different distributions, we first recall how it is possible to introduce similar statistics for arbitrary distributions.

We consider variables $x_1$, $x_2$, ..., and suppose that these have arbitrary distributions $P_1(x_1)dx_1$, etc., on the relevant null hypotheses. Then the statistic f, defined by

$$f = F(x_1, x_2, ...), \qquad (A.1)$$

is distributed according to the probability distribution function $P_F(f)$, where

$$P_F(f)df = \left[ \iint ... dx_1 dx_2 ... P_1(x_1) P_2(x_2) ... \delta(f - F(x_1, x_2, ...)) \right] df. \qquad (A.2)$$

If each variable has a normal distribution with mean zero and standard deviation unity, and if we adopt the function

$$F(x_1, x_2, ...) = x_1^2 + x_2^2 + ..., \qquad (A.3)$$

f has the familiar chi-square distribution (Bartoszynki & Niewiadomska-Bugaj 1996, p. 757; Rice 1988, p. 168).

If we apply the operation (1.3) to the statistic defined by equation (A.2),

$$\Gamma = -\ln\left( \int_f^\infty df\, P_F(f) \right), \qquad (A.4)$$

we arrive at a statistic that is distributed exponentially.

APPENDIX B

If we wish to combine information from n power spectra, the combination that is closest to the chi-square statistic is the sum of the powers, which we write as

$$Z = S_1 + S_2 + ... + S_n. \tag{B.1}$$

We see from (A.2) that the probability distribution function is given by

$$P_{C,n}(Z)dZ = \left[\int_0^\infty \int_0^\infty ...dx_1 dx_2...e^{-x_1-x_2-...}\delta(Z - x_1 - x_2 - ...)\right]dZ. \tag{B.2}$$

This is found to be

$$P_{C,n}(Z) = \tfrac{1}{(n-1)!} Z^{n-1} e^{-Z}, \tag{B.3}$$

that is known as the "gamma distribution" (Bartoszynki & Niewiadomska-Bugaj 1996, p. 384; Rice 1988, p. 168).

Equation (B.3) may be verified by induction. We see from (B.2) that

$$P_{C,n+1}(Z)dZ = \int_0^\infty dx_1 ... \int_0^\infty dx_{n+1} e^{-x_1-...-x_{n+1}} \delta(Z - x_1 - ... - x_{n+1})dZ, \tag{B.4}$$

so that

$$P_{C,n+1}(Z) = \int_0^\infty dx\, e^{-x} P_{C,n}(Z-x), \tag{B.5}$$

which leads to

$$P_{C,n+1}(Z) = \tfrac{1}{n!} Z^n e^{-Z}. \tag{B.6}$$

confirming equation (B.3).

We may now carry out the operation of equation (A.4) in order to obtain a statistic that is distributed exponentially. We write

$$G_n(Z) = -\ln\left(\int_Z^\infty dz\, P_{C,n}(z)\right), \tag{B.7}$$

and so arrive at the following formula for the "combined power statistic:"

$$G_n(Z) = Z - \ln\left(1 + Z + \tfrac{1}{2}Z^2 + \ldots + \tfrac{1}{(n-1)!}Z^{n-1}\right). \tag{B.8}$$

## APPENDIX C

We may wish to determine the frequencies for which the minimum power among n power spectra has the maximum value. If we write

$$U(x_1, x_2, \ldots) = Min(x_1, x_2, \ldots), \tag{C.1}$$

we see from (A.2) that the probability distribution function for U is given by

$$P_{M,n}(U)dU = \left[\iint \ldots dx_1 dx_2 \ldots P_1(x)P_2(x_2)\ldots \delta(U - Min(x_1, x_2, \ldots))\right]dU. \tag{C.2}$$

We now verify by induction that

$$P_{M,n}(U) = n e^{-nU}. \tag{C.3}$$

We see from (C.1) and (C.2) that

$$P_{M,n+1}(V)dV = \left[\iint dU\, dx\, P_{M,n}(U) e^{-x} \delta(V - Min(U, x))\right]dV. \tag{C.4}$$

We may separate the integral over x into two parts: $x = 0$ *to* $U$ and $x = U$ *to* $\infty$. Hence, using (C.3), we arrive at

$$P_{M,n+1}(U) = (n+1)e^{-(n+1)U}, \tag{C.5}$$

so that (C.3) has been confirmed.

Following (A.4), we now write

$$K_n(U) = -\ln\left(\int_u^\infty dy\, P_{M,n}(y)\right), \tag{C.6}$$

to obtain a statistic, that we call the "minimum power statistic," that is distributed exponentially. This has the simple form

$$K_n(U) = nU. \tag{C.7}$$

## APPENDIX D

If we introduce the notation

$$Y = S_1 S_2, \tag{D.1}$$

we see from (A.2) that the probability distribution function for Y is given by

$$P_{J,2}(Y)dY = \int_0^\infty \int_0^\infty du\, dv\, e^{-u-v}\delta(Y-uv)dY, \tag{D.2}$$

i.e., by

$$P_{J,2}(Y) = \int_0^\infty \frac{du}{u} e^{-u-Y/u}. \tag{D.3}$$

We may use equation (A.4) to transform this statistic to one, that we call the "joint power statistic," that is distributed exponentially:

$$J_2(Y) = -\ln\left(\int_Y^\infty dy\, P_{J,2}(y)\right), \tag{D.4}$$

which becomes

$$J_2 = -\ln\left(\int_0^\infty du\, e^{-u-Y/u}\right). \tag{D.5}$$

This is found to be expressible as

$$J_2 = -\ln\left(2Y^{1/2} K_1\left(2Y^{1/2}\right)\right), \tag{D.6}$$

where $K_1$ is the Bessel function of the second kind.

We see that it would be more convenient to introduce the notation

$$X = (S_1 S_2)^{1/2}, \tag{D.7}$$

so that (D.6) becomes

$$J_2 = -\ln(2X K_1(2X)). \tag{D.8}$$

The leading terms of the asymptotic expression for $J_2$ are as follows:

$$J_2 \to 2X - \tfrac{1}{2}\ln(\pi) - \tfrac{1}{2}\ln(X) \text{ as } X \to \infty. \tag{D.9}$$

## APPENDIX E

We now consider joint power statistics of higher orders. If

$$Y = S_1...S_n ,\tag{E.1}$$

we see from (A.2) that the probability distribution function for $Y$ is given by

$$P_{J,n}(Y)dY = \left[\int_0^\infty dx_1 ... \int_0^\infty dx_n e^{-x_1-...-x_n} \delta(Y - x_1...x_n)\right] dY .\tag{E.2}$$

By examining the corresponding formula for $P_{n+1}(Y)$, we find that

$$P_{J,n+1}(Y) = \int_0^\infty dx\, e^{-x} P_{J,n}(Y/x) .\tag{E.3}$$

In terms of the cumulative distribution functions,

$$C_{J,n}(Y) = \int_Y^\infty dy\, P_{J,n}(y) ,\tag{E.4}$$

the iterative relation (E.3) becomes

$$C_{J,n+1}(Y) = \int_0^\infty dx\, e^{-x} C_{J,n}(Y/x) .\tag{E.5}$$

We have used this iterative relation to calculate joint power statistics of the third and fourth orders. Application of the iterative relation (E.5) gives the following expressions for $C_{J,3}$ and $C_{J,4}$:-

$$C_{J,3}(X) = \int_0^\infty dx\, e^{-x} 2 \frac{X^{3/2}}{x^{1/2}} K_1\left[2X^{3/2}/x^{1/2}\right]\tag{E.6}$$

and

$$C_{J,4}(X) = \int_0^\infty dx \int_0^\infty dx'\, e^{-(x+x')} \frac{2X^2}{(xx')^{1/2}} K_1\left[2X^2/(xx')^{1/2}\right].\tag{E.7}$$

The joint power statistics of third and fourth orders are then given by

$$J_3(X) = -\ln[C_{J,3}(X)], \quad J_4(X) = -\ln[C_{J,4}(X)] . \tag{E.8}$$

## APPENDIX F

We now develop approximate formulae for the joint spectrum statistics of second, third and fourth orders. After trial and error, we were led to consider the form

$$J = \frac{aX^2}{b+X} . \tag{F.1}$$

We could have attempted a nonlinear process for determining the optimum values of the coefficients a and b. However, it was simpler and adequate to rewrite (F.1) as

$$X^2 a - Jb = XJ . \tag{F.2}$$

Given a set of values X and J, determined analytically for J2 or by computation for J3 and J4, we could then determine the coefficients a and b by the least-squares process. The results of these calculations are given in Section 4.

## APPENDIX G

Note that one may obtain approximate expressions for the higher-order joint power statistics by combining the results of Appendix E and Appendix F. Beginning with an approximate formula for $J_2, J_3,$ or $J_4$, as developed in Appendix F, we may compute the cumulative distribution function by using (1.2). We may now use (E.5) to obtain an estimate of the cumulative distribution function for the statistic of next higher order, from which we may obtain the statistic itself by means of (1.3). We may then seek an approximate formula for this statistic by the procedure of Appendix F. This procedure may be repeated for as long as the approximations prove to be acceptable.

REFERENCES

Bai, T.A. 2003, ApJ 591, 406

Bahcall, J.N., & Press, W.H. 1991, ApJ, 370, 730

Bartoszynki, R., & Niewiadomska-Bugaj, M. 1996, Probability and Statistical Inference (New York:Wiley)

Caldwell, D.O., & Sturrock, P.A. 2003, Phys. Rev. Letters (submitted).

Cleveland, B.T., et al. 1995, Proc. Nucl. Phys. B (Proc. Suppl), 38, 47

Cleveland, B.T., et al. 1998, ApJ, 496, 505

Davis Jr, R., & Cox, A.N. 1991, Solar Interior and Atmosphere (eds. A.N. Cox, W.C. Livingston & M.S. Matthews;    Tucson: Univ. Arizona Press), 51

Lande, K., et al. 1992, AIP Conf. Proc., No. 243. Particles and Nuclear Physics, ed. W.T.H. van Oers (New York: American Institute of Physics), 1122

Oppenheim, A.V., Schafer, R.W., & Buck, J.R. 1999, Discrete-Time Signal Processing (2nd ed., New York: Prentice Hall)

Press, W.H., Teukolsky, S.A., Vetterling, W.T., & Flannery, B.P. 1992, Numerical Recipes (Cambridge University Press), p. 570

Rice, J.A. 1988, Mathematical Statistics and Data Analysis (Wadsworth & Brooks, Pacific Grove, California)

Rieger, E., Share. G.H., Forrest, D.J., Kanbach, G., Reppin, C., & Chupp, E.L. 1984, Nature 312, 623

Saio, H. 1982, ApJ, 256, 717

Scargle, J.D. 1982, ApJ, 263, 835

Sturrock, P.A. 2003a, hep-ph/0304106

Sturrock, P.A. 2003b, BAAS. 35. 822

Sturrock, P.A. 2003c, hep-ph/0304148

Sturrock, P.A., Walther, G., & Wheatland, M.S. 1997, ApJ, 491, 409

Sturrock, P.A., Walther, G., & Wheatland, M.S. 1998, ApJ, 507, 978

Sturrock, P.A., & Weber. M.A. 2002a, "Multi-Wavelength Observations of Coronal Structure and Dynamics -- Yohkoh 10th Anniversary Meeting", COSPAR Colloquia Series, P.C.H. Martens and D. Cauffman (eds.), p. 323

Sturrock, P.A., & Weber, M.A. 2002b, ApJ, 565, 1366

Walther, G. 1999, ApJ, 513, 990

Wolff, C.L. 2002, ApJ (Letters), 580, L181

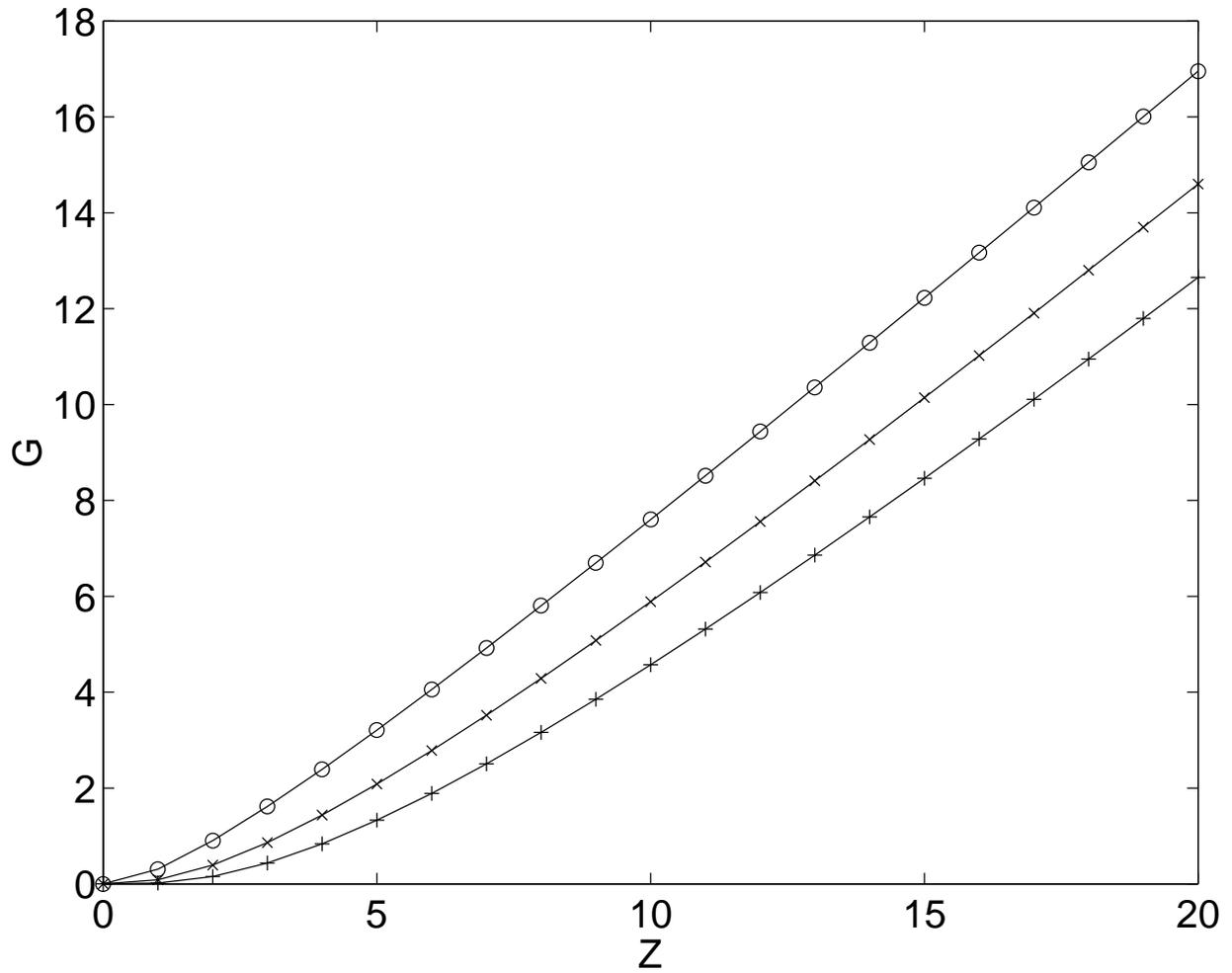

Figure 1. Combined spectrum statistics of second, third and fourth order: G2 ('o'), G3 ('x'), and G4 ('+').

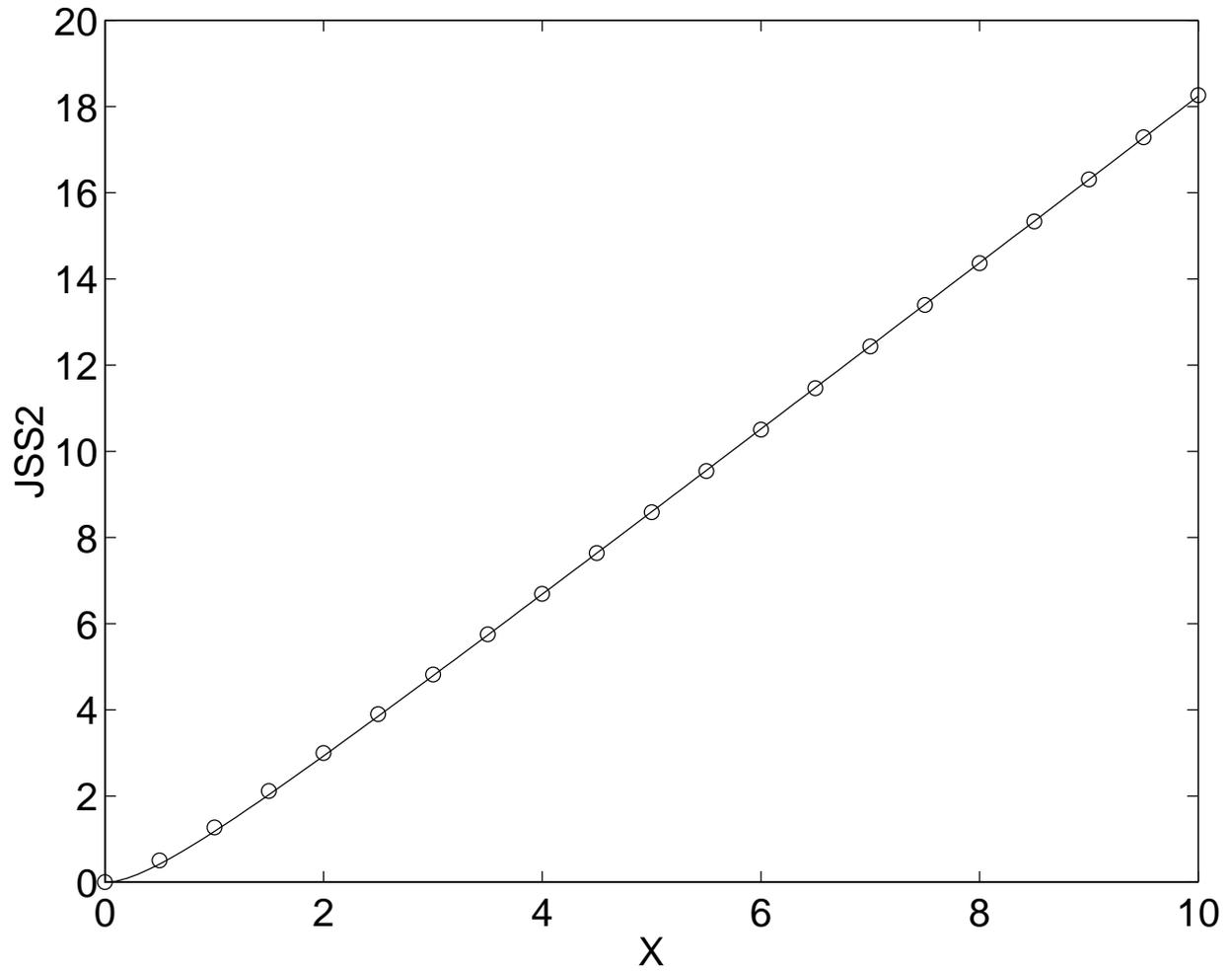

Figure 2. Joint spectrum statistic of second order as given by the approximate expression (4.3) and (in circles) as calculated from the analytical expression (4.2).

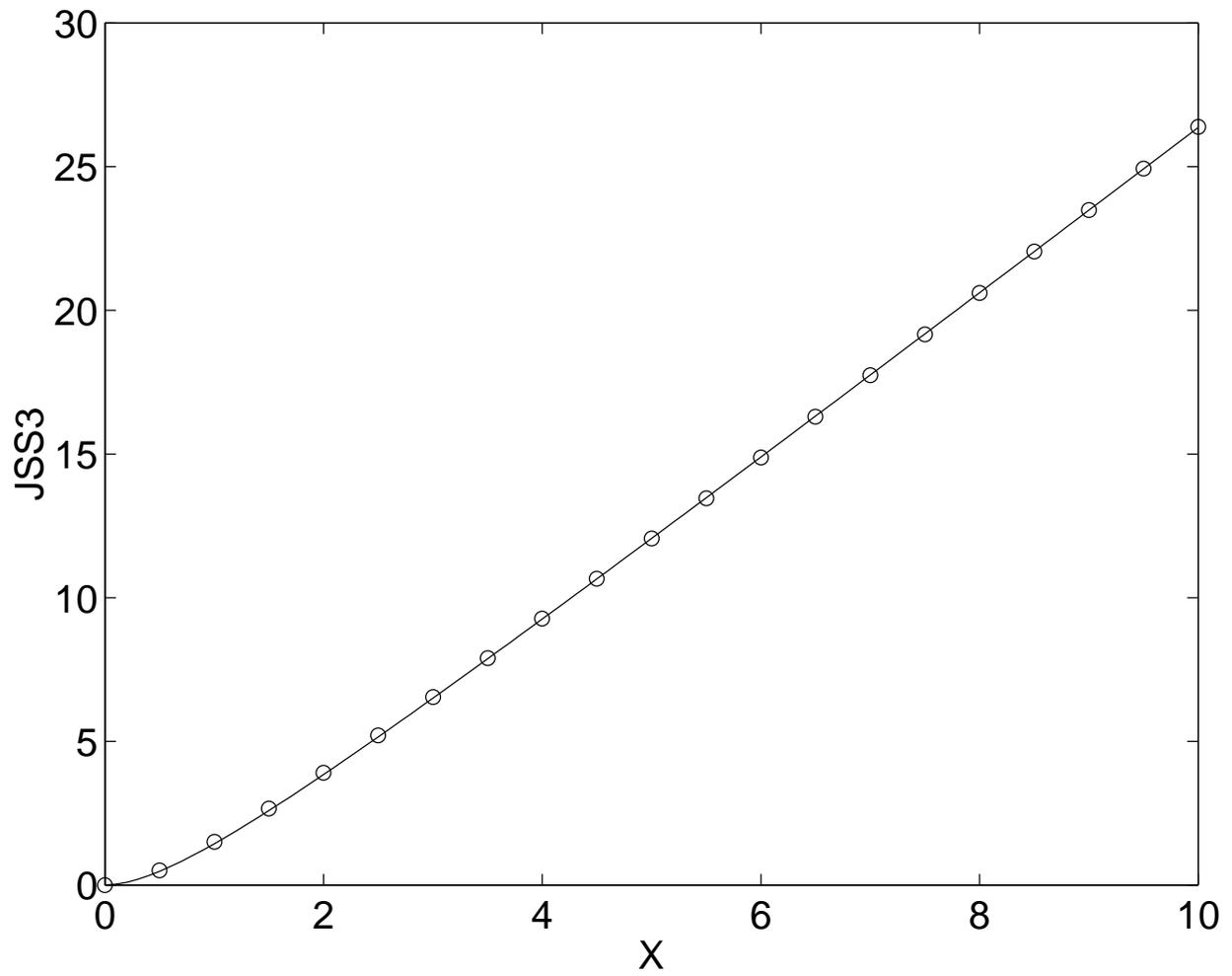

Figure 3. Joint spectrum statistic of third order as given by the approximate expression (4.5) ) and (in circles) as given in Table 1, calculated by the method outlined in Appendix E.

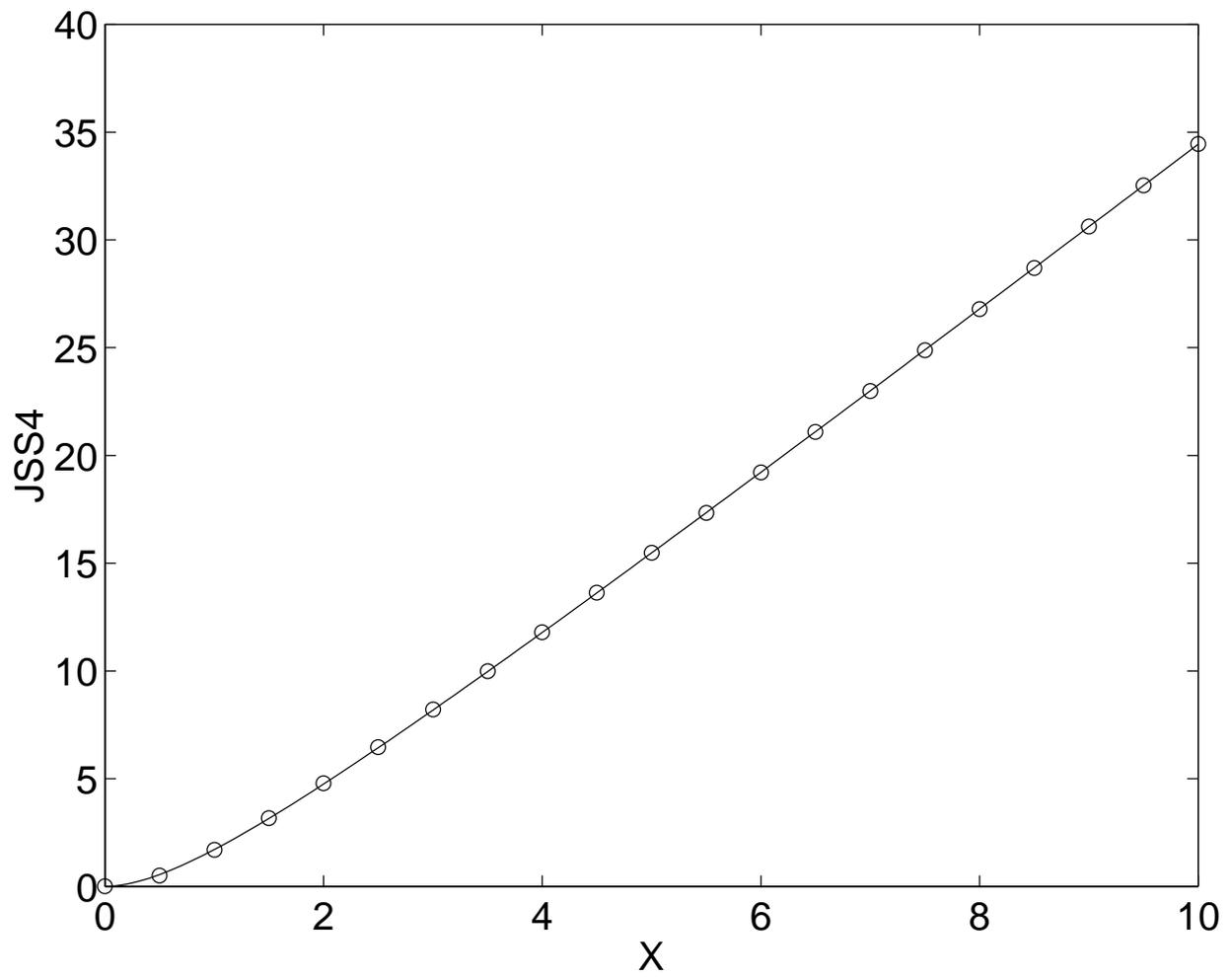

Figure 4. Joint spectrum statistic of fourth order as given by the approximate expression (4.6) and (in circles) as given in Table 2, calculated by the method outlined in Appendix E.

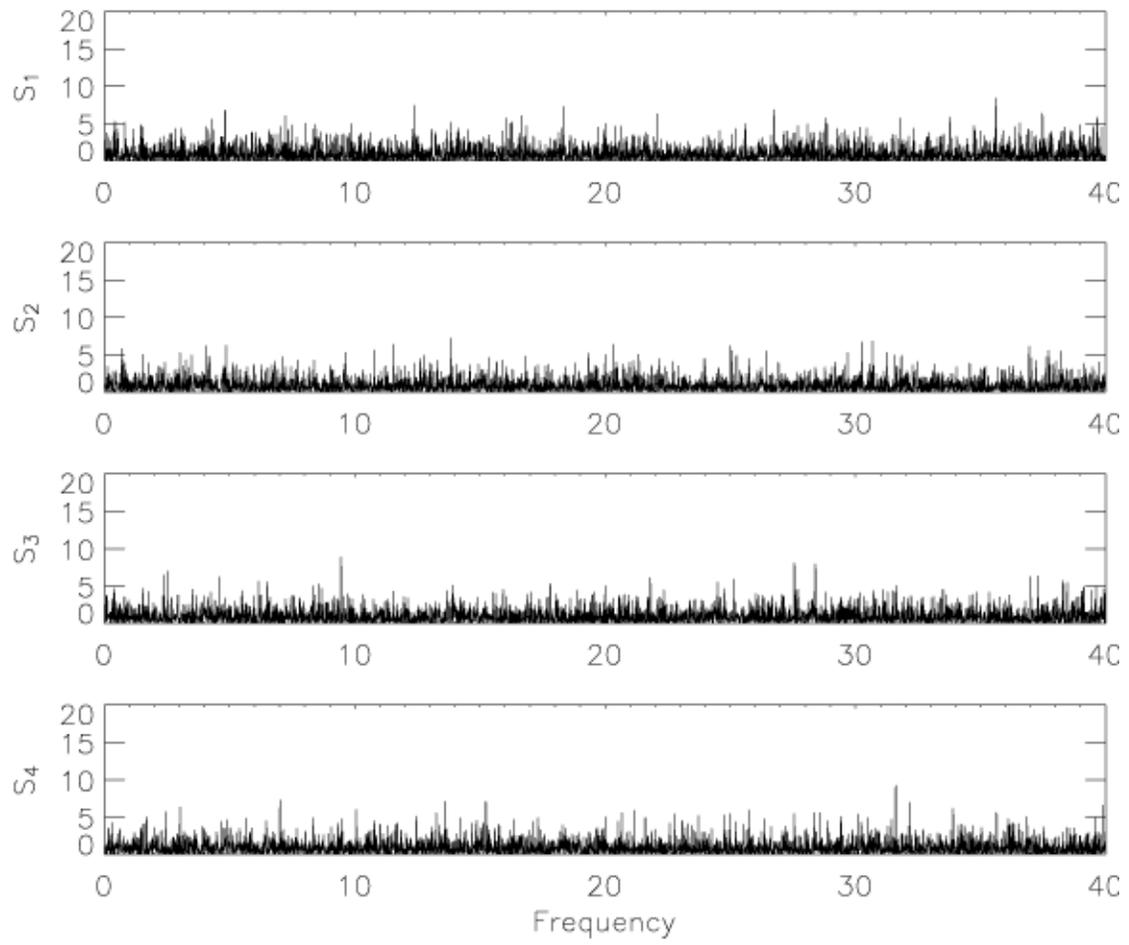

Figure 5. Four synthetic spectra, each with a signal of power 5 at $\nu = 20$.

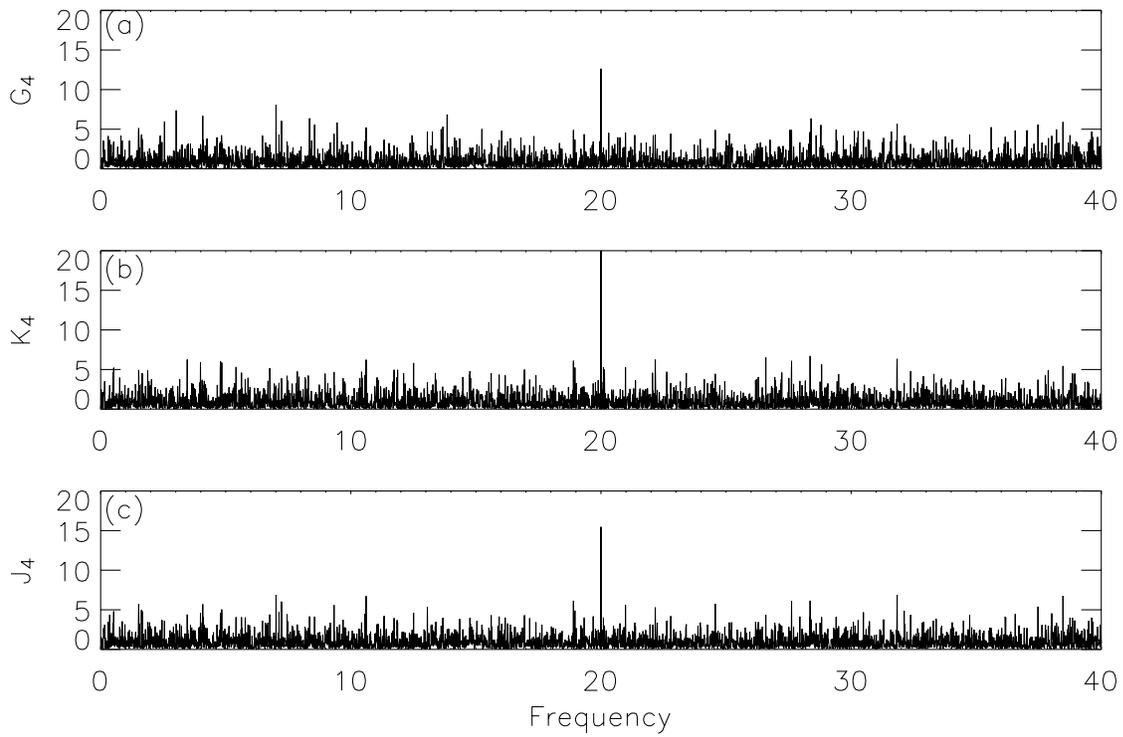

Figure 6. The combined power statistic, minimum power statistic, and joint power statistic, formed from the four synthetic spectra shown in Figure 5.

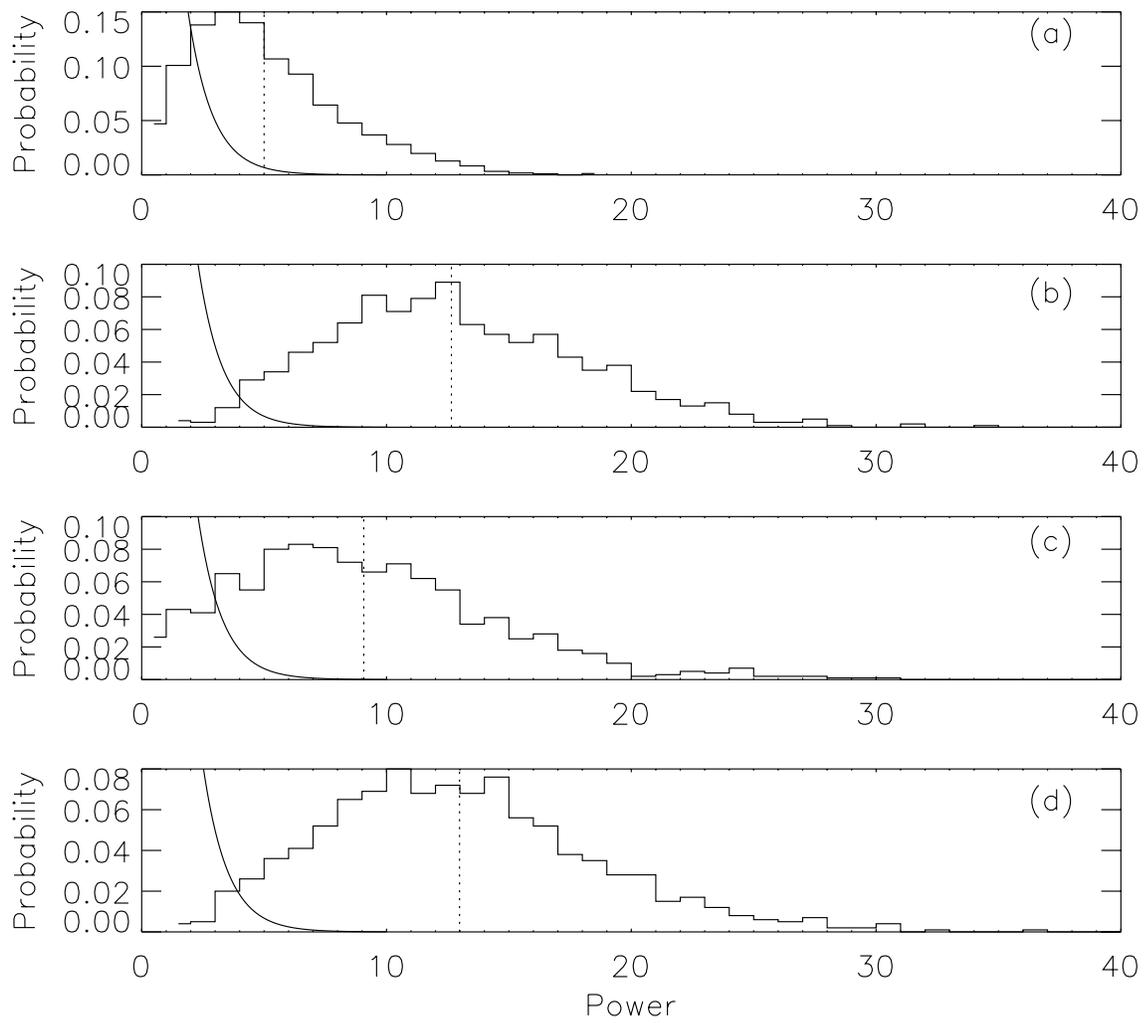

Figure 7. Distribution of statistics at the signal frequency for 1000 sets of four spectra produced from synthetic time series: (a) power spectrum; (b) combined power statistic; (c) minimum power statistic; and (d) joint power statistic.

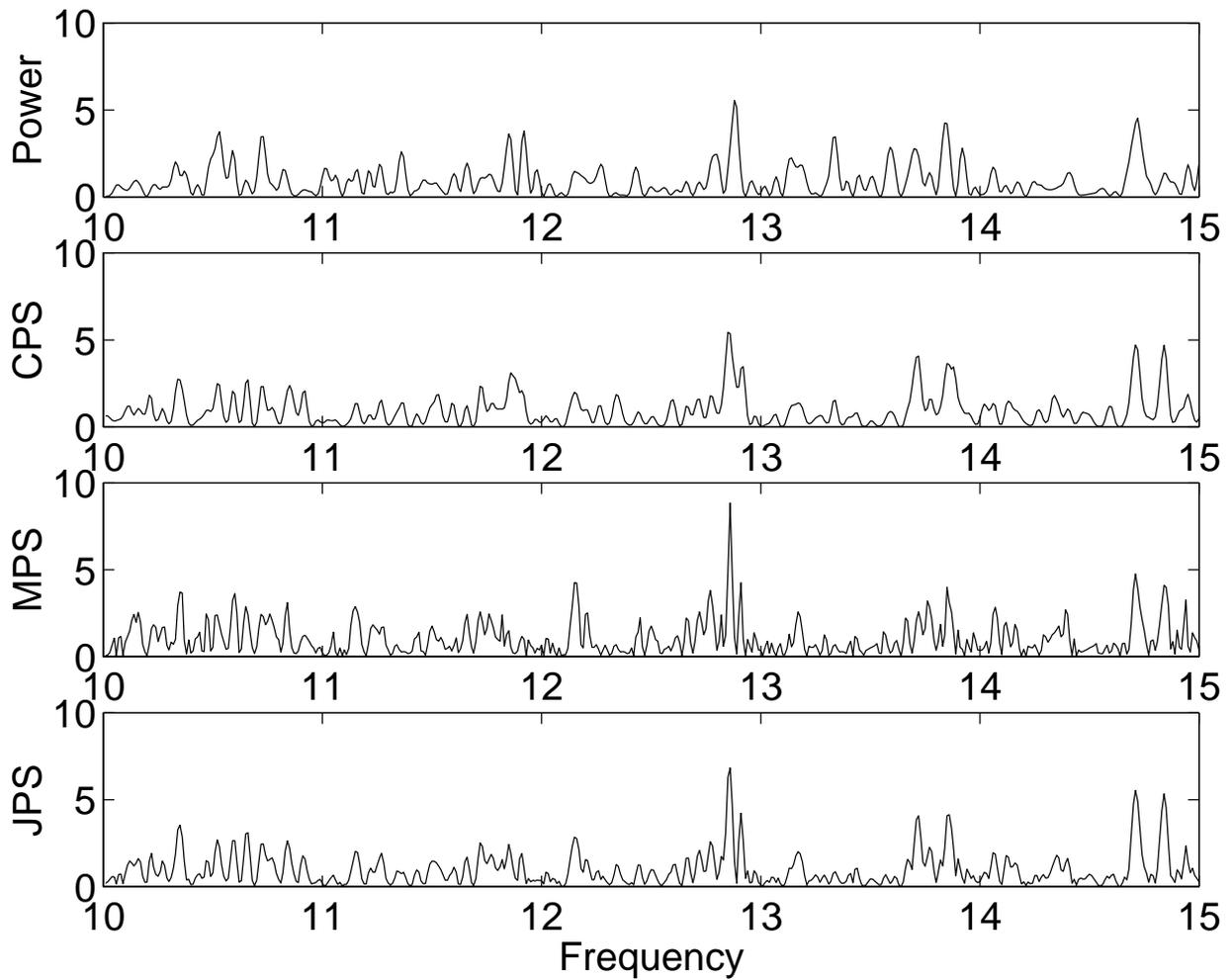

Figure 8. Power spectrum of Homestake data and the combined spectrum statistic of third order calculated according to equation (6.1). Note that the combined spectrum statistic clarifies the case that the peak at 12.88 is accompanied by sidebands near 11.88 and 13.88.